\documentclass{article}
\usepackage[utf8]{inputenc}

\usepackage[usenames]{color}
\usepackage{epsfig}
\usepackage{graphics}
\usepackage{amsmath}
\usepackage{amssymb}
\usepackage{multirow}
\usepackage{cite}
\usepackage{array}
\usepackage{pslatex} 
\usepackage{url}
\usepackage{caption}
\usepackage{lineno}
\usepackage{graphicx}  
\usepackage{setspace}
\usepackage{tikz}
\usepackage{hhline}
\usepackage{mathtools}
\usepackage[a4paper, total={7in, 10in}]{geometry}
\usepackage{subfigure}
\usepackage{stfloats}
\usepackage{colortbl}
\usepackage{flushend}
\definecolor{light-gray}{gray}{0.9}

\usepackage{algorithm}
\usepackage[noend]{algpseudocode}

\begin{document}
\title{CovidRhythm: A Deep Learning Model for Passive Prediction of Covid-19 using Biobehavioral Rhythms Derived from Wearable Physiological Data}

\author{Atifa Sarwar*, Emmanuel O. Agu \\
Worcester Polytechnic Institute, Worcester, MA, USA \\
*Corresponding Author: asarwar@wpi.edu}
\date{}
\maketitle

\begin{abstract}
\textit{Goal:}  To investigate whether a deep learning model can detect Covid-19 from disruptions in the human body's physiological (heart rate) and rest-activity rhythms (rhythmic dysregulation) caused by the SARS-CoV-2 virus. \textit{Methods:} We propose \textit{CovidRhythm}, a novel Gated Recurrent Unit (GRU) Network with Multi-Head Self-Attention (MHSA) that combines sensor and rhythmic features extracted from heart rate and activity (steps) data gathered passively using  consumer-grade smart wearable to predict Covid-19. A total of 39 features were extracted (standard deviation, mean, min/max/avg length of sedentary and active bouts) from wearable sensor data. Biobehavioral rhythms were modeled using nine parameters (mesor, amplitude, acrophase, and intra-daily variability). These features were then input to CovidRhythm for predicting Covid-19 in the incubation phase (one day before biological symptoms manifest).  \textit{Results:} A combination of sensor and biobehavioral rhythm features achieved the highest AUC-ROC of 0.79 [Sensitivity = 0.69, Specificity=0.89, F$_{0.1}$ = 0.76], outperforming prior approaches in discriminating Covid-positive patients from healthy controls using 24 hours of historical wearable physiological. Rhythmic features were the most predictive of Covid-19 infection when utilized either alone or in conjunction with sensor features. Sensor features predicted healthy subjects best. Circadian rest-activity rhythms that combine 24h activity and sleep information were the most disrupted. \textit{Conclusions}: CovidRhythm demonstrates that biobehavioral rhythms derived from consumer-grade wearable data can facilitate timely Covid-19 detection. To the best of our knowledge, our work is the first to detect Covid-19 using deep learning and biobehavioral rhythms features derived from consumer-grade wearable data.

\end{abstract}

\section{INTRODUCTION}
\label{sec:introduction}

Covid-19 is an infectious disease caused by the Severe Acute Respiratory Syndrome Coronavirus 2 (SARS-CoV-2) \cite{WHO}. Covid-19 was unknown before its first outbreak in Wuhan, China, in late December 2019 and was declared a global pandemic by the World Health Organization (WHO) on March 11, 2020\cite{whopanademic}. 
Covid-19 vaccinations were introduced in early 2021, significantly reduced adverse outcomes including non-Intensive Care Unit (ICU) hospitalizations and deaths by 63.5\% and 69.5\% respectively in United States \cite{moghadas2021impact}. However, vaccination faces several challenges as an intervention. Some subjects are  unwilling to receive Covid-19 vaccines for various reasons including its perceived safety \cite{karlsson2021fearing} and religious beliefs \cite{olagoke2021intention}. Due to its fast transmission rate, the Covid-19 pandemic highlighted the need for early detection of viral infections to facilitate timely public health interventions, clinical management and disease spread. 

Most living organisms have an internal biological clock and bio-rhythms that regulate their physiological functions, including performance, sleep, rest-activity cycles, and mood. The biological clock is controlled by the human brain and 
ensures synchrony between an organism's external and internal environment, which is critical to the well-being and survival of an organism\cite{reinberg2022concepts}. A lack of synchrony can have significant health ramifications~\cite{vitaterna2001overview}. Based on the periodicity with which these bio-rhythms reoccur, they can be categorized as ultradian ($<$ 24hrs), circadian ($=$ 24hrs), or infradian ($>$ 24hrs). Persistent disruption of a human being's rhythms is associated with various health conditions including cardiovascular disorders \cite{takeda2011circadian}, psychiatric and neurodegenerative disease \cite{wulff2010sleep}, obesity \cite{laermans2016chronobesity}, and depression \cite{difrancesco2019sleep}. The link between circadian rhythms and viral diseases has previously been established by \cite{anderson2020melatonin} \cite{sengupta2019circadian}. Recent studies have found that viruses, including SARS-CoV-2, disrupt circadian regulation in order to enhance their replication \cite{ray2020circadian}. Viruses invade host cells bound to angiotensin-converting enzyme II (ACE), which displays circadian expression patterns in tissues from which it is released. Over-expression of the ACE2 enzyme increases the replication of SARS-CoV-2, resulting in symptom severity \cite{wu2020new}.

Actigraphy devices have been widely employed in healthcare and in chronobiology, facilitating the understanding of circadian rhythms, sleep-wake cycles or rest-activity cycles and their relationship with various health conditions~\cite{yang2022rest}~\cite{diaz2022sleep}~\cite{roh2022associations}. Widely-owned mobile devices and smart wearables can now passively monitor physiological data such as heart rate, physical activity, sleep, and body temperature, which has paved the way for passive, continuous monitoring and study of circadian rhythms and ailment-specific disruptions without requiring specialized research-grade equipment \cite{lee2017comparison}. \textit{It is important to note that while ailment-specific tests are often invasive and require visits to the clinics, the disruptions in circadian rhythms caused by such ailments can be measured by consumer grade wearables, facilitating low burden monitoring outside the clinic.}

\begin{figure*}[t]
\centering
\subfigure[]{
\includegraphics[width=7cm]{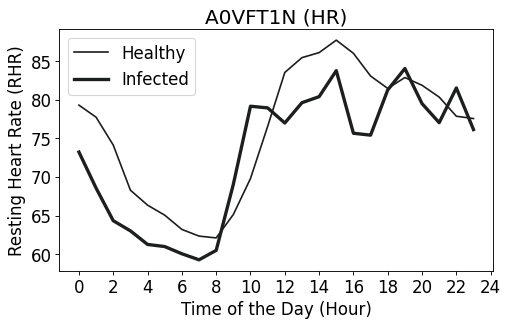}
\label{fig:hrinfectedvshealthy}}
\subfigure[]{
\includegraphics[width=7cm]{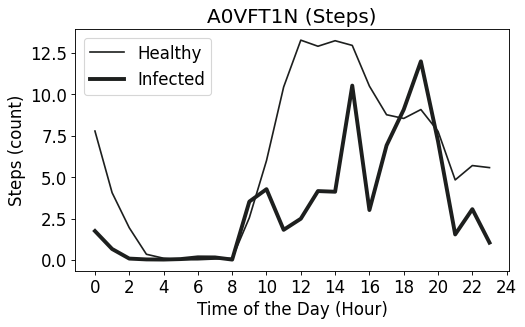}
\label{fig:stepshealthyvsinfected}}
\caption{Daily rhythm of RHR for Subject A0VFT1N when (a) Healthy (b) Covid-Positive. Average profile of RHR and steps over all days (Healthy vs Infected) is depicted.}
\label{fig:hrsteps}
\end{figure*}

\begin{figure*}[t]
\centering
\subfigure[]{
\includegraphics[width=7cm]{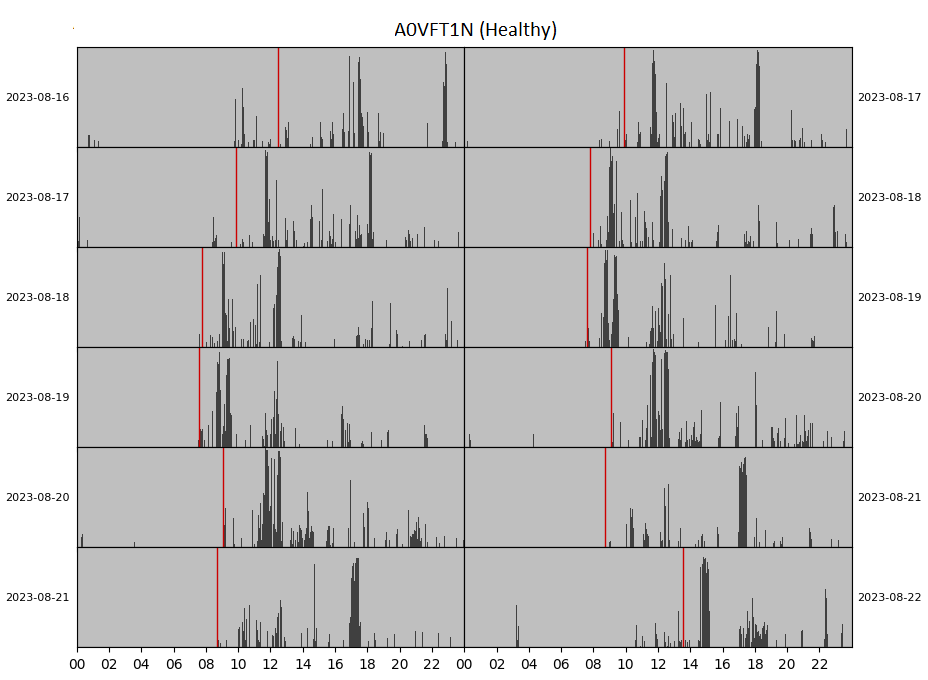}
\label{fig:rarHealthy}}
\subfigure[]{
\includegraphics[width=7cm]{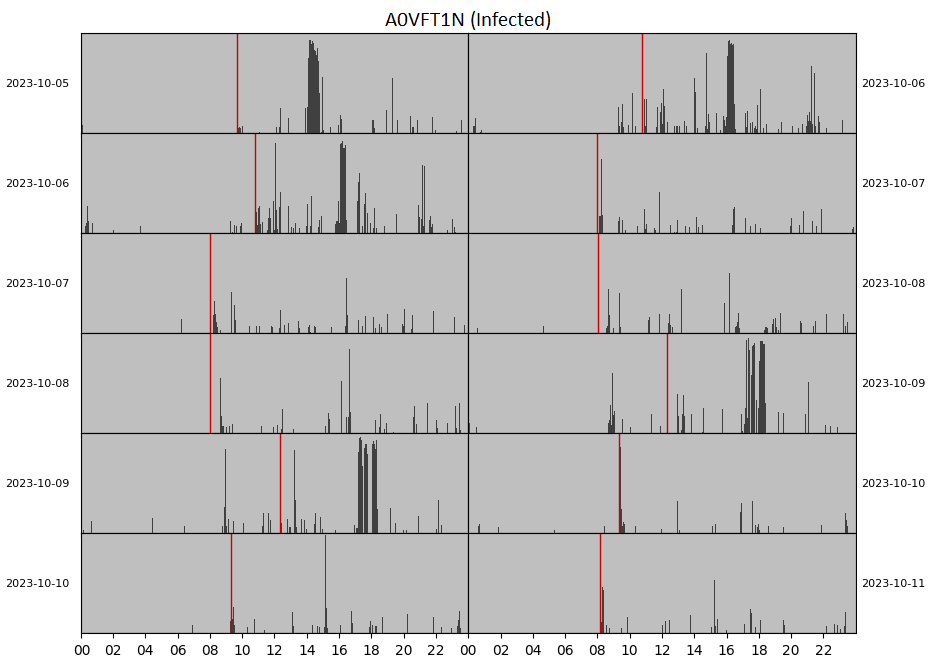}
\label{fig:rarInfected}}
\caption{Double plot of Circadian Rest Activity Rhythm for Subject A0VFT1N when (a) Healthy (b) Covid-19 Positive. Day 1 and Day 2 are plotted consecutively (48h) and beneath one another to enhance visualizations of the patterns.}
\label{fig:raractrogram}
\end{figure*}

The overarching goal of this paper is to explore the detectability of Covid-19 infection using deep learning approaches using biobehavioral rhythm features derived from wearable data. Prior research and studies about Covid-19 assessment using circadian rhythms derived from wearable data have mainly focused on: I) Establishing statistical relationships between Covid-19 and circadian rhythms \cite{lee2021changes} \cite{merikanto2022disturbances}, II) Exploiting circadian rhythms to improve the clinical management of Covid-19 \cite{ray2020covid}, and III) Investigating the impact of lifestyle changes (lockdown, social distancing) during Covid-19 pandemic on circadian rhythms \cite{salehinejad2021circadian} \cite{bertrand2022sleep} \cite{carrigan2020sleep}. Only Mayer et al. \cite{mayer2022consumer} previously studied the impact of heart rate circadian rhythms on machine learning prediction of Covid-19, finding that circadian phase uncertainty is a strong predictive attribute for pre-detection of Covid-19 infection. However, their study only focused on heart rate circadian rhythms and trained the model on a balanced dataset by focusing on parameter classification between specific windows. Moreover, their training dataset did not contain participants who were healthy throughout the study duration (healthy controls)~\cite{nestor2021dear}. Our work is different in that we aim to detect Covid-19 infection using biobehavioral rhythm features derived from heart rate and rest-activity (sleep-wake) data, training the model on an imbalanced dataset with unequal numbers of healthy and infected subjects. To mitigate balance class distributions, minority class samples were replicated, an approach demonstrated to be effective by prior studies. Deep learning was also employed, which typically outperforms traditional machine learning for most tasks as long as adequate data is available. To the best of our knowledge, till date, no prior work has passively detected Covid-19 by utilizing data from consumer-grade smart wearables to derive   \textit{biobehavioral rhythms} (includes cyclic physiological (heart rate), psychological (mood), environmental (weather), and rest-activity patterns \cite{doryab2019modeling}).


\begin{figure*}[t]
\centering
    \includegraphics[width=14cm]{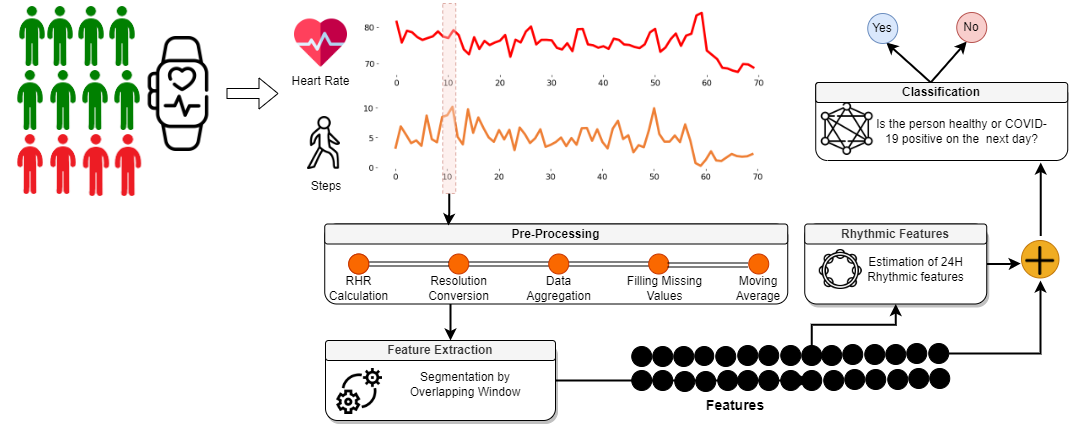}
    \caption{Overview of our approach for Covid-19 classification using a fusion of Sensor and Rhythmic Features}
    \label{fig:approach}
\end{figure*}


Figure \ref{fig:hrsteps} shows the daily average Resting Heart Rate (RHR) and steps profile of a subject when healthy (thin line) and Covid-positive (bold line). During nighttime (12:00 AM to 8:00 AM), the subject's RHR decreases when infected vs. healthy. A sudden disruption is apparent for the rest of the day, highlighting the misalignment in the RHR rhythm. Similarly, Figure \ref{fig:stepshealthyvsinfected} shows a decrease in the amount of physical activity. Even though the subject aimed to perform physical activity, a drastic decrease can be observed in the step counts over the day. These findings were further established from actograms of the rest-activity patterns (See Figure \ref{fig:raractrogram}). When healthy, the subject displayed a distinct and regular day-night pattern, with more activity being displayed during the day than night (see Figure \ref{fig:rarHealthy}). However, when infected, this distinction and regularity were less apparent. Patients also exhibited some daytime napping and decreased physical activity. These observations provide preliminary qualitative evidence that Covid-19 infection disrupts the physiological (heart rate) and rest-activity rhythms. The rest of the paper focuses on determining whether these disruptions can be leveraged to predict Covid-19 infection using a neural networks model.
  
 We propose \textit{CovidRhythm}, a novel neural network model,  which builds on prior work that has demonstrated qualitative and associative relationships between bio-rhythms and Covid-19. \textit{CovidRhythm} analyzes hourly heart rates and step data, gathered passively using a Fitbit wearable. A total of 39 sensor features were extracted including those that captured resting heart rate (standard deviation and slope) and physical activity (total number of active bouts, length of sedentary and active bouts, total steps when physically active). Additionally, biobehavioral rhythmic features were extracted including mesor, amplitude, M10, L5. These features were then input to \textit{CovidRhythm} to identify each subject as healthy or infected during the incubation period (one day before symptom onset). CovidRhythm utilizes a Gated Recurrent Unit (GRU) to learn temporal patterns in sensor features, followed by a Multi-Head Self Attention (MHSA) module that combines  information in different parts of an input sequence, yielding richer and more predictive representations. \textit{CovidRhythm} then concatenates the resultant output with rhythmic features, which is utilized for the final prediction. In our evaluation using five-fold cross-validation with subject-wise splitting, \textit{CovidRhythm}  achieved an average Area Under the Curve (AUC) of 0.79, with 0.69 sensitivity and 0.89 specificity.  Rhythmic features were the most predictive of Covid-19 infection when utilized either alone or in conjunction with sensor features. Sensor features predicted healthy subjects best.  
 Covid-19 affected circadian rhythms, which repeat every 24 hours, the most. Rest-activity rhythms, which combines activity and sleep information, were the most disrupted, reflecting the sedentary behaviors and irregular sleep patterns of subjects once infected. The rest of the paper is as follows: Section \ref{methodology} describes our methodology including an overview of the dataset, pre-processing techniques, feature extraction, and classification. Section \ref{result} evaluates \textit{CovidRhythm}, and Section \ref{discussion} highlights important findings, future work and limitations of the study. Section \ref{conclusion} concludes the paper.
%
%
%
%

\section{MATERIALS AND METHODS }
\label{methodology}
This section describes \textit{CovidRhythm}, our proposed deep MHSA-based GRU neural network for capturing disruptions in biobehavioral rhythms for the purpose of detecting Covid-19 infection. Fig \ref{fig:approach} provides a general overview of our proposed approach.

\subsection {Covid-19 Dataset}
We analyzed a dataset previously gathered by Mishra et al. that explored early detection of Covid-19 using physiological data gathered from a smartwatch~\cite{mishra2020early}. The study enrolled 5,262 participants, out of which 4642 reported wearing a smartwatch, with 3325 using a Fitbit, 984 using an Apple device, and 428 using a Garmin. Since most enrolled participants used Fitbit, the study gathered data from only Fitbit users and had 70 healthy, 25 Covid positives, and 11 non-COVID-illness individuals. For creating a labeled dataset, we randomly selected a 24-hour interval starting from 00:00 to 23:59 from healthy subjects and labeled them as healthy. In contrast, one day before symptom onset date was selected as an infected sample. In this study, our dataset was restricted  to only healthy and Covid-19 subjects. Since the number of healthy and infected subjects was highly imbalanced, the infected samples were replicated to the number of healthy samples~\cite{liu2021fitbeat} to guide the detection model weighted in favor of the minority class.

\begin{algorithm}[h] 
\caption{Calculating Resting Heart Rate (RHR)}
\label{alg:loop}
\begin{algorithmic}[1]
\Require{Heart Rate: $h_{1} \dots h_{T}$ and  Step : $s_{1} \dots s_{T}$ } 
\Ensure{$RHR$ (Resting heart rate)}
\Statex
\Function{Resting heart rate}{$hr[\;], s[\;]$}
   \State {$RHR$ $\gets$ {$null$}}
    \State {$T$ $\gets$ {$length(heartrate)$}}
    \State {$pre\_hr$ $\gets$ {$0$}}
    \For{$k \gets 1$ to $T$} 
    \State {$Sum$ $\gets$ {$0$}}
       \For{$m \gets k$ to $k+5$}
        \State {$Sum$ $\gets$ {$Sum +s[m]$}}
        \EndFor 
            \If{$\textit{Sum} == \textit{0}$}
                \State {$RHR[k] = h[k]$ 
                \State {$pre\_hr= h[k]$ }} 
            \Else {
                \State {$RHR[k] = pre\_hr$ }}
            \EndIf
        \EndFor
\State \Return {$RHR$}
  \EndFunction
\end{algorithmic}
\end{algorithm}

\subsection{Pre-Processing} 
First, Resting Heart Rate (RHR) was computed, defined as the user's heart rate when they were inactive for a sufficiently long time~\cite{radin2020harnessing} (See Algorithm \ref{alg:loop}). RHR is considered to be the true representation of cardiac health as it is not influenced by any external factor such as physical activity. All the days with missing values exceeding 10\% of overall readings on that day were discarded. As Fitbit Charge 3 \& 4 need about 1-2 hours to charge \cite{fitbit}, the threshold of 10\% was set to ensure that missing values existed only because the devices were removed for charging. For the remaining days, missing values were filled using linear interpolation given as: $f(x) = f(x_0) + \frac{f(x_1) - f(x_0)}{x_1 - x_0}(x - x_0)$. To synchronize multiple data streams, we resampled the RHR to a one-minute resolution, and then aggregated  using step data based on the RHR timestamp. Finally, we used the simple moving average $SMA_n = \sum_{t =n-T}^{n-1} \frac{x_t}{T}$ to smooth the time series data over 24 hour period.

\textit{Segmentation: } The raw sensor data was segmented into 60-minute windows with a 50\% overlap of consecutive windows. The impact of various overlapping sizes on performance will be studied in section \ref{result}.

\subsection{Feature Extraction} \label{sec:feature}
CovidRhythm analyzes two types of features. I) Sensor II) Biobehavioral Rhythmic.  This section provides a general overview of the features extracted with their mathematical expressions described in Table \ref{tab:features}.

\subsubsection{Sensor Features}\label{statistical}
Statistical and behavioral features of RHR, and steps, motivated by prior work by Doryab et al.~\cite{doryab2018extraction}\cite{doryab2019modeling} and Singh et al. \cite{singh2020machine}, were extracted from raw Fitbit sensor data.

\begin{itemize}
    \item \textbf{Resting Heart Rate (RHR):} 21 features including i) Mean ii) Median iii) Variance iv) Standard deviation v) Interquartile Range (IQR) vi) Range vii) Skewness viii) Kurtosis ix) Second Momentum x) Entropy xi) Slope xii) Max., Min., and average positive change xiii) Max., Min., and average negative change xiv) Max., Min., and average absolute change, and xv) No change
    \item \textbf{Steps: } 18 features including i) Total no. of steps ii) Average no. of steps iii) Standard Deviation iv) Variance v) Entropy vi) Max. no. of steps taken in any 5-minute interval vii) Total no. of active bouts viii) Total no. of sedentary bouts ix) Max., Min. and average length of active or sedentary bout x) Min., Max., and average no. of steps over all active bouts and xi) Slope.
\end{itemize}

\subsubsection{Biobehavioral Rhythmic Features}
We used the Cosinor \cite{halberg1967circadian}  for rhythm analysis due to its ability to: I) Model both equidistant and non-equidistant data, and II) Provide means to estimate the rhythmicity parameter that we used further in our deep learning model. Based on the trigonometry regression model, Cosinor presents a fundamental method that obtains an estimate of the Midline Statistic of Rhythm (MESOR), the amplitude, a measure of phase (acrophase) for the chosen period using cosinor curve fitting. When the period is known, the model is defined as:     
    \[y(t) = \sum_{i=1}^{N}(A_{i,1} * sin(\frac{t}{P/i}.2\pi) + A_{i,2} * cos(\frac{t}{P/i}.2\pi))\\ + M + e(t)\]

where $t$ corresponds to the time points to be observed within the time series, $N$ is the number of components, $M$ is MESOR, $P$ is the observed period, $e(t)$ is the error term, and $A_{i,1}$ and $A_{i,2}$ are the parameters of the model. If the period is unknown, different feasible ranges can be evaluated to identify the optimal period. We employed CosinorPY \cite{movskon2020cosinorpy}, a publicly available Python implementation of cosinor-based methods for rhythmicity detection. For each sensor feature, nine biobehavioral rhythmic feature for 24h, 48h, and 96h were extracted, given as:

\begin{itemize}
    \item \textit{MESOR} is the midline of the oscillatory function and refers to the mean daily activity.
    \item \textit{Acrophase} is the time when the rhythm reaches its maximal value for the first time in the cycle.
    \item \textit{Amplitude} refers to the  maximum value a rhythm can rise above or below the mesor.
    \item \textit{Relative Amplitude} is defined as the amplitude of the circadian rhythm model divided by mesor.
    \item \textit{Multi-Scale Entropy (MSE):} calculates the pearson's sample entropy, a measure of the complexity of time series data, at different time scales.
    \item \textit{Most Active 10 hour (M10) and Least Active 5 hour (L5):} M10 refers to the mean activity during the most active 10h, while L5 represents the mean activity during the least active 5h. 
    \item \textit{Rest activity Relative Amplitude (RA):} is defined as the difference between a person's daily activity (M10) and nocturnal activity (L5).
    \item \textit{Intra-daily Variability (IV):} is a measure of circadian disturbance, namely, capturing how activity level shifts between two consecutive hours.
    \end{itemize}
 \begin{table*}[t]
  \caption{List of Sensor and Biobehavioral Rhythmic Features extracted}
  \label{tab:features}
  \begin{center}
    \renewcommand{\arraystretch}{1.5}
  \begin{tabular}{p{3cm}p{4cm}p{3cm}p{4cm}}
    \hline
    \rowcolor{light-gray}
    \textbf{Feature} & \textbf{Description} & \textbf{Feature} & \textbf{Description}\\
    \hline
    Mean & \(\displaystyle \frac{\sum_{n=1}^{N} n }{N} \) &  Standard Deviation & \(\displaystyle \sqrt{ \frac{\sum_{i=1}^{N} (x_i - \mu)^2 }{N}} \)  \\ 
    \hline
    Median & \(\displaystyle \begin{cases}
    X[\frac{n}{2}],& \text{if n is even}\\
    \frac{X[\frac{n-1}{2}] + X[\frac{n+1}{2}]}{2},              & \text{otherwise}
\end{cases} \) &
 Variance &  \(\displaystyle \frac{\sum_{i=1}^{N} (x_i - \mu)^2 }{N-1} \)   \\ 
    \hline
    IQR &  \(\displaystyle Quartile3 - Quartile1\) &    Range &  \(\displaystyle Max(X) - Min (X)\)   \\ 
    \hline
    Second Momentum &\(\displaystyle \frac{1}{n}\sum_{i=1}^{n} (x_i - \mu) ^ 2 \) &  Entropy & \(\displaystyle - \sum_{}^{} Xlog(X) \)  \\ 
    \hline
    Skewness & \(\displaystyle \frac{\frac{1}{N}\sum_{i=1}^{N}(x_i - \mu)^3}{ (\frac{1}{N-1}\sum_{i=1}^{N}(x_i - \mu)^2)^\frac{3}{2}}\) &
    Kurtosis & \(\displaystyle \frac{\frac{1}{N}\sum_{i=1}^{N}(x_i - \mu)^4}{ (\frac{1}{N-1}\sum_{i=1}^{N}(x_i - \mu)^2)^2} - 3\)\\
    \hline
     Slope & \(\displaystyle Y=aX + b\) \newline where b is the slope over time & Bout & Continuous period when a certain characteristic is exhibited \\
      \hline
   Sedentary Bout & $<$ than 10 steps during a 5-minute interval &    Active Bout & $>$ than 10 steps during a 5-minute interval \\
     \hline
     Relative Amplitude & $\frac{amplitude}{mesor}$ & Multi-Scale Entropy (MSE) & $-\ln \frac{U^{m+1}}{U^m}$ \\
     \hline 
     Rest Activity Relative Amplitude & $\frac{M10 - L5}{M10 + L5}$ & Intra-daily Variability (IV) & $ \frac{N \sum_{i=2}^{N}(x_i - x_{i-1})^2}{(N-1) \sum_{i=1}^{N}(x_i - \mu)^2}$\\
     \hline
\end{tabular}
 \end{center}
\end{table*}
\subsection{Feature Selection}\label{featureselection}
As previously described, we extracted 39 sensor features from heart rate and steps data collected using smart wearables. For each sensor feature, we further computed nine rhythmic features, resulting in total of $ 39 \times 9 = 351$ features which is quite large compared to the relatively few data samples for training. We thus used Mutual Information (MI, also referred as Information Gain) \cite{witten2005practical} to select the most relevant features for predicting the target variable.  MI quantifies the amount of information contained by features with respect to the target variable as $I(X,Y) = H(X) - H(X|Y)$ where where $X$ and $Y$ are two random variables, $I(X,Y)$ is the MI for $X$ and $Y$, and $H(X)$ is the entropy of $X$. For both sensor and rhythmic features, we selected ten features (discussed in section \ref{result}) with the highest information gain, leaving the following features: 

\subsubsection{Rhythmic Features}\label{lb:rythmic10}
\begin{itemize}
    \item \textbf{Resting Heart Rate:} M10 for RHR Skewness, M10 for RHR Max Positive Change
    \item \textbf{Step:} Intra-daily Variability for Total Steps taken, Average Steps taken, and Standard Deviation, MSE for Variance, L5 for Max Sedentary Bout length, Relative Amplitude for Average Sedentary Bout length, Mesor and L5 for Slope.
\end{itemize}
\begin{figure}[t]
\centering
    \includegraphics[width=9cm]{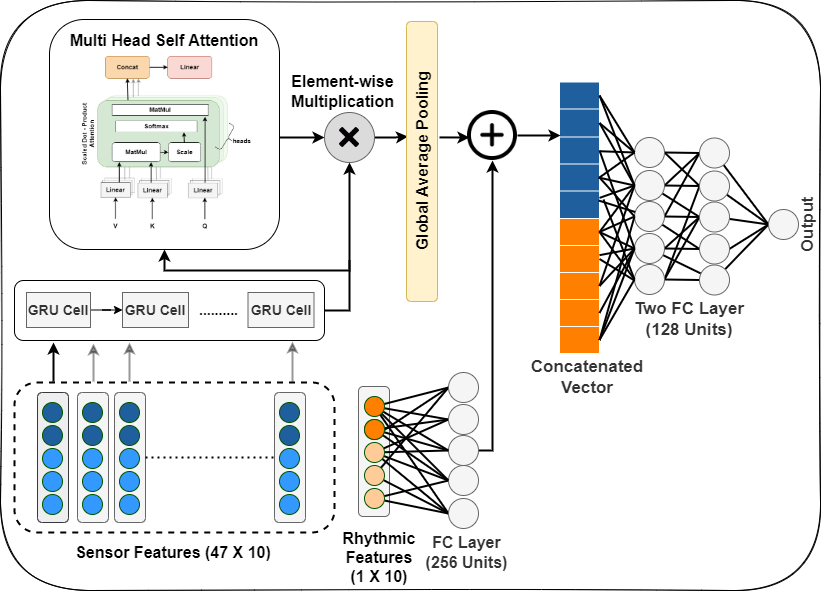}
    \caption{Architecture of \textit{CovidRhythm}}
    \label{fig:model}
\end{figure}

\subsubsection{Sensor Features}
\begin{itemize}
    \item \textbf{Resting Heart Rate:} Mean, Median, Slope, Min Positive Change, Max Positive Change, Min Negative Change, Average Negative Change, Min Absolute Change
    \item \textbf{Step:} Max Steps taken in any 5-min interval, and Max Steps taken in active bout.
\end{itemize}

\subsection{Model Architecture}
An overview of \textit{CovidRhythm}  architecture is shown in Figure \ref{fig:model}. The model analyzes input sensor features of a whole day segmented into hourly intervals with 50\% overlap between adjacent segments, yielding a tensor of shape 47 X 10, where ten refers to features per timestamp. The input is passed from a single Gated Recurrent Unit (GRU) \cite{cho2014learning} layer, in order to learn temporal dependencies between timestamps. A GRU was selected as they I) Learn rare content better such as abnormal physiological signs that occur for very short time periods and have values that have very slight differences from baseline values. II) Can outperform other temporal neural networks architectures (LSTM, RNN) in terms of speed of convergence, and generalization \cite{chung2014empirical}. In time series data, not every data point has the same importance. Especially in terms of vital signs, medical practitioners generally examine only critical measurements rather than reviewing all points. Multi-Head Self-Attention (MHSA) \cite{vaswani2017attention} learns different weighted representations of data inputs, i.e., the most important dependencies to achieve high attention scores, suppressing irrelevant information. Consequently, an MHSA is employed right after the GRU layer in order to learn the most important datapoints for making the final prediction. The resultant output is concatenated with the rhythmic features (1 X 10), and passed through two fully connected layers, followed by dropout (0.25), to estimate the joint representation of sensor and rhythmic features. Finally, the sigmoid is used to produce the final output, i.e., whether the subject is infected or healthy.

\section{Experimental Setup}\label{setup}
\subsection{Validation Protocols}
To rigorously evaluate \textit{CovidRhythm}  performance, we utilized $5$-fold cross-validation, where no overlap exists between training and testing segments at each fold. For evaluation, we employed AUC-ROC, sensitivity, specificity, and F$_{\beta}$ score defined as:
\begin{itemize}
    \item Sensitivity measures the proportion of diseased subjects that were correctly identified, given as: $\frac{TP}{TP + FN}$
     \item Specificity measures the proportion of healthy subjects that were correctly identified, given as: $ \frac{TN}{TN + FP}$
     \item  AUC-ROC is a probability curve that plots \textit{TPR} and \textit{FPR} at various thresholds and evaluates how well the model distinguishes between classes.
     \item F$_{\beta}$ score, a generalization of F$_{1}$score with an additional parameter $\beta$, is defined as the weighted harmonic mean of precision and recall, given as :$F_{\beta} = \frac{((1 + \beta^2) \times Precision \times Recall)}{\beta^2 \times Precision  + Recall}$
     \end{itemize}

\subsection{Normalization and Hyperparameter Tuning}

Each instance $z$ of the train and test dataset was normalized using $z = \frac{x-\mu}{s}$
where $\mu$ is the mean of training samples and $s$ is standard deviation. We implemented this normalization using the \textit{StandardScaler} function in the \textit{sklearn} library, while the classifier is implemented using Keras library. To achieve the best results, we explored and tuned a number of hyper-parameters using grid search. Table \ref{tab:hyperparamter} shows the optimum values of these hyperparameters along with the values that were originally explored.
\begin{table}[t]
  \caption{Training Hyperparameters}
  \label{tab:hyperparamter}
  \renewcommand{\arraystretch}{1.5}
  \begin{center}
  \begin{tabular}{p{4cm}p{4.5cm}p{7cm}}
    \hline
    \rowcolor{light-gray}
    \textbf{Hyperparameter} & \textbf{Explored Values} & \textbf{Optimum Value}\\
    \hline
     Recurrent Depth & 1, 2, 4 & 1 \\
    \hline
     Heads & 0, 1, 2, 4, 8 & 2  \\
     \hline
    Batch Size & 8, 16, 32 & 8 \\
    \hline
     GRU Layer Size & 256, 512 & 256  \\
    \hline
    Loss Function & Binary Cross Entropy, Focal Loss & Binary Cross Entropy  \\
    \hline
    Optimizer & Adam, SGD, RMS Prop & RMS Prop  \\
    \hline
    Fully Connected Layers & 1, 2, 3, 4 & 1 (Rhythms Network), 2 (Joint Network) \\
    \hline
    Neurons in Hidden Layer & 128, 256, 512 & 256 (Rhythms Network), 128 (Joint Network) \\
    \hline
    Dropout & 0.25, 0.5 & 0.25 \\
    \hline
    Learning Rate & 0.001 & Exponentially decayed with 90\% decay rate \\
    \hline
    Epochs & 100 & Early Stopping used to avoid overfitting \\
    \hline
\end{tabular}
 \end{center}
\end{table}    

\section{Results}\label{result}
\subsection{Comparing \textit{CovidRhythm}  to Baselines}
\textit{CovidRhythm} was compared to eight baseline models that have previously been utilized to predict infection from physiological signs. \textit{CovidRhythm}  outperformed all baselines with an AUC-ROC of 0.79, a sensitivity of 0.69, a specificity of 0.89, and an F$_{0.1}$ score of 0.76 when trained with an overlap of 50\% among consecutive windows (Table \ref{tab:results}.) We observed that overlapping sizes are associated with an improved performance, which in our case increases by 0.05 for AUC-ROC and sensitivity, 0.06 for specificity, whereas a significant increase was seen in F$_{0.1}$, which increases from 0.64 to 0.76 when the model is trained with no vs. 50\% of overlapping. For some metric, the performance of a few learning algorithms were comparable to that of \textit{CovidRhythm}. For instance, Gradient Boosting achieved an AUC-ROC of 0.78, sensitivity of 0.78, and specificity of 0.79 when trained with an overlap of 25\% between two consecutive windows. However, when comparing the precision, recall, and F$_{0.1}$ of all the algorithms, we found that although Gradient Boosting performed close to \textit{CovidRhythm's}, a significant difference was observed in the F$_{0.1}$ score, with the former achieving an F$_{0.1}$ of 0.6. In comparison, \textit{CovidRhythm} achieved the highest F$_{0.1}$ score of 0.76.

\begin{table*}[!h]
 \caption{CovidRhythm vs. Baselines (Sens. refers to Sensitivity, Spec. to Specificity and AUC to AUC-ROC)}
  \label{tab:results}
  \begin{tabular}{p{1.8cm}p{0.75cm}p{0.85cm}p{0.85cm}p{1cm}p{0.85cm}p{0.85cm}p{0.85cm}p{1cm}p{0.85cm}p{0.85cm}p{0.85cm}p{0.9cm}}
\hline
 &
\multicolumn{4}{c}{\textbf{o = No Overlap}} &
\multicolumn{4}{c}{\textbf{o = 25\% Overlap}} &
\multicolumn{4}{c}{\textbf{o = 50\% Overlap}} \\
\hline

\textbf{Algo} & \textbf{Sens.} & \textbf{Spec.} &\textbf{AUC}& \textbf{F$_{0.1}$} & \textbf{Sens.} & \textbf{Spec.} & \textbf{AUC} & \textbf{F$_{0.1}$}& \textbf{Sens.} & \textbf{Spec.} &\textbf{AUC}& \textbf{F$_{0.1}$} \\
\hline
Naive Bayes& 0.48$_{\pm{0.1}}$ & 0.53$_{\pm{0.2}}$ & 0.56$_{\pm{0.1}}$ & 0.3$_{\pm{0.1}}$ & 0.52$_{\pm{0.1}}$ & 0.52$_{\pm{0.3}}$ & 0.5$_{\pm{0.1}}$ & 0.35$_{\pm{0.2}}$ & 0.44$_{\pm{0.1}}$ & 0.51$_{\pm{0.2}}$ &0.49$_{\pm{0.2}}$&0.27$_{\pm{0.1}}$\\
\hline
Random Forest  & 0.27$_{\pm{0.3}}$ & 0.97$_{\pm{0.1}}$ & 0.66$_{\pm{0.1}}$ & 0.6$_{\pm{0.3}}$ &0.23$_{\pm{0.3}}$ & 0.97$_{\pm{0.1}}$ & 0.69$_{\pm{0.1}}$& 0.5$_{\pm{0.4}}$& 0.22$_{\pm{0.2}}$ & 0.98$_{\pm{0.1}}$ & 0.77$_{\pm{0.1}}$ &0.68$_{\pm{0.4}}$\\
\hline
SVM   & 0.36$_{\pm{0.3}}$ & 0.82$_{\pm{0.2}}$ & 0.59$_{\pm{0.1}}$ &0.4$_{\pm{0.3}}$ & 0.27$_{\pm{0.1}}$ & 0.85$_{\pm{0.1}}$ & 0.56$_{\pm{0.1}}$ & 0.4$_{\pm{0.1}}$ &0.36$_{\pm{0.3}}$ & 0.90$_{\pm{0.1}}$ & 0.63$_{\pm{0.1}}$ &0.4$_{\pm{0.1}}$ \\
\hline
Gradient Boosting  & 0.43$_{\pm{0.2}}$ & 0.87$_{\pm{0.2}}$ & 0.7$_{\pm{0.2}}$ &0.65$_{\pm{0.3}}$ & 0.78$_{\pm{0.1}}$  &  0.79$_{\pm{0.1}}$  & 0.78$_{\pm{0.1}}$  & 0.6$_{\pm{0.2}}$ &0.35$_{\pm{0.2}}$ & 0.82$_{\pm{0.2}}$ & 0.65$_{\pm{0.1}}$ &0.38$_{\pm{0.3}}$\\
\hline
AdaBoost   & 0.53$_{\pm{0.3}}$ & 0.85$_{\pm{0.1}}$ & 0.69$_{\pm{0.2}}$ & 0.5$_{\pm{0.3}}$ &0.43$_{\pm{0.2}}$ & 0.87$_{\pm{0.1}}$ & 0.69$_{\pm{0.2}}$ &0.61$_{\pm{0.2}}$ & 0.59$_{\pm{0.4}}$ & 0.87$_{\pm{0.1}}$ & \textbf{0.87}$_{\pm{0.1}}$ &0.58$_{\pm{0.4}}$\\
\hline
CNN  & 0.65$_{\pm{0.1}}$ & 0.85$_{\pm{0.2}}$ & 0.75$_{\pm{0.1}}$ &0.7$_{\pm{0.2}}$ & 0.66$_{\pm{0.2}}$ & 0.82$_{\pm{0.2}}$ & 0.74$_{\pm{0.1}}$ &0.65$_{\pm{0.2}}$ & 0.66$_{\pm{0.2}}$ & 0.82$_{\pm{0.2}}$ & 0.74$_{\pm{0.1}}$ &0.65$_{\pm{0.2}}$ \\
\hline
LSTM  & 0.56$_{\pm{0.3}}$  & 0.8$_{\pm{0.2}}$ & 0.68$_{\pm{0.2}}$ &0.57$_{\pm{0.3}}$ & 0.57$_{\pm{0.2}}$ & 0.85$_{\pm{0.1}}$ & 0.71$_{\pm{0.1}}$ &0.6$_{\pm{0.3}}$ & 0.52$_{\pm{0.2}}$ & 0.82$_{\pm{0.2}}$ & 0.68$_{\pm{0.3}}$ &0.58$_{\pm{0.4}}$\\
\hline
CNN + LSTM   & 0.69$_{\pm{0.1}}$ & 0.80$_{\pm{0.2}}$ & 0.75$_{\pm{0.2}}$ &0.67$_{\pm{0.3}}$ & 0.56$_{\pm{0.3}}$ & 0.83$_{\pm{0.2}}$ & 0.70$_{\pm{0.2}}$ &0.6$_{\pm{0.3}}$ & 0.69$_{\pm{0.1}}$ & 0.80$_{\pm{0.2}}$ & 0.75$_{\pm{0.2}}$ & 0.64$_{\pm{0.3}}$\\
\hline
\textbf{CovidRhythm} & 0.64$_{\pm{0.2}}$ & 0.83$_{\pm{0.2}}$ & 0.74$_{\pm{0.2}}$ &0.66$_{\pm{0.3}}$ &  0.64$_{\pm{0.2}}$ & 0.88$_{\pm{0.1}}$ & 0.76$_{\pm{0.1}}$ &0.55$_{\pm{0.3}}$ & \textbf{0.69}$_{\pm{0.1}}$ & \textbf{0.89}$_{\pm{0.1}}$ & 0.79$_{\pm{0.1}}$ & \textbf{0.76}$_{\pm{0.2}}$\\
 \hline
\end{tabular}
\end{table*}

\subsection{Feature Importance}
\begin{figure*}[t]
\centering
\includegraphics[width=0.85\textwidth]{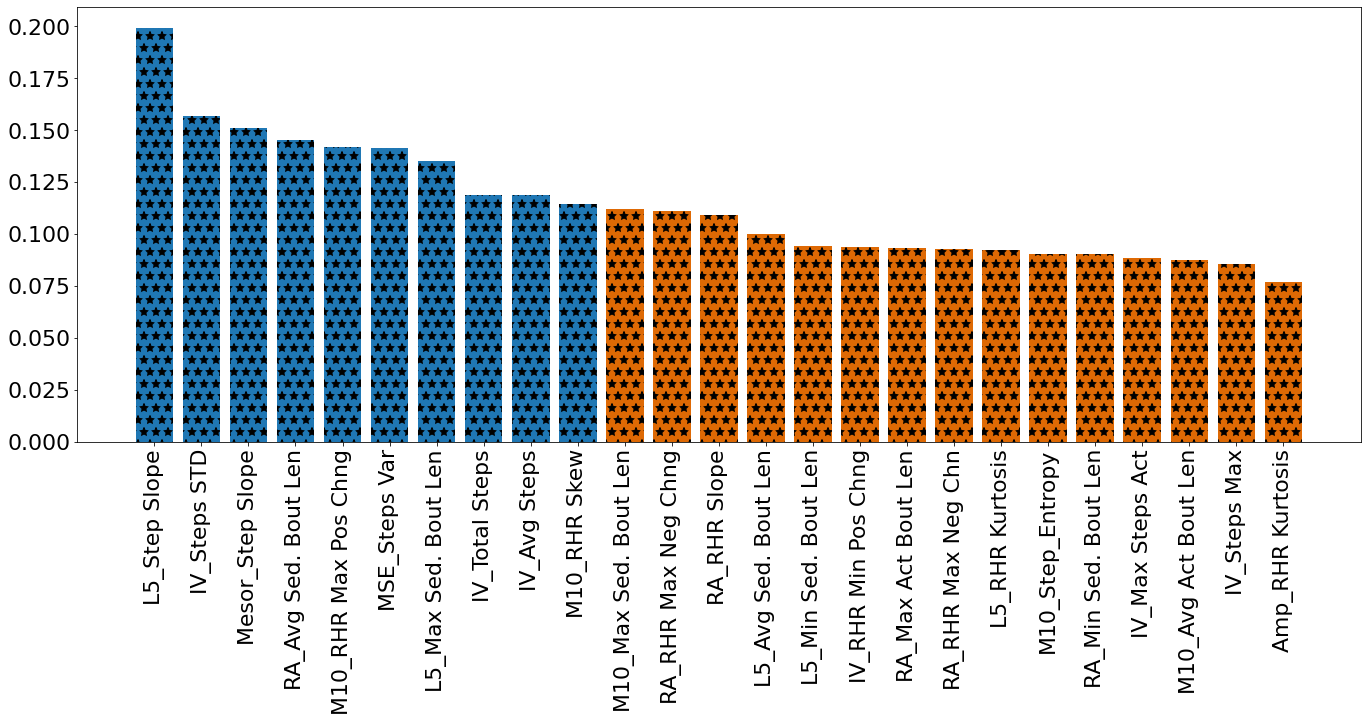}
\caption{Top 25 Rhythmic Features. The top 10 Features (Shown in Blue) are utilized by \textit{CovidRhythm}  for Prediction. RA=Relative Amplitude, Sed.=Sedentary, Act.=Active, and Chng=Change}
\label{fig:rhythmfeatureImp}
\end{figure*}

Figure \ref{fig:rhythmfeatureImp} highlights the top 25 most important rhythmic features, computed by Mutual Information (MI).  Disruption in rest-activity rhythms was the most prominent, as 20 of the top 25 features were related to the disruption in rest-activity rhythms. In contrast, only five relate to physiological rhythms (resting heart rate). The findings implied that although Covid-19 infection affects both physiological and rest-activity rhythms, the latter are more predictive of the infection.

\subsection{Dimension Reduction}
\begin{figure*}[t]
\centering
\includegraphics[width=0.85\textwidth]{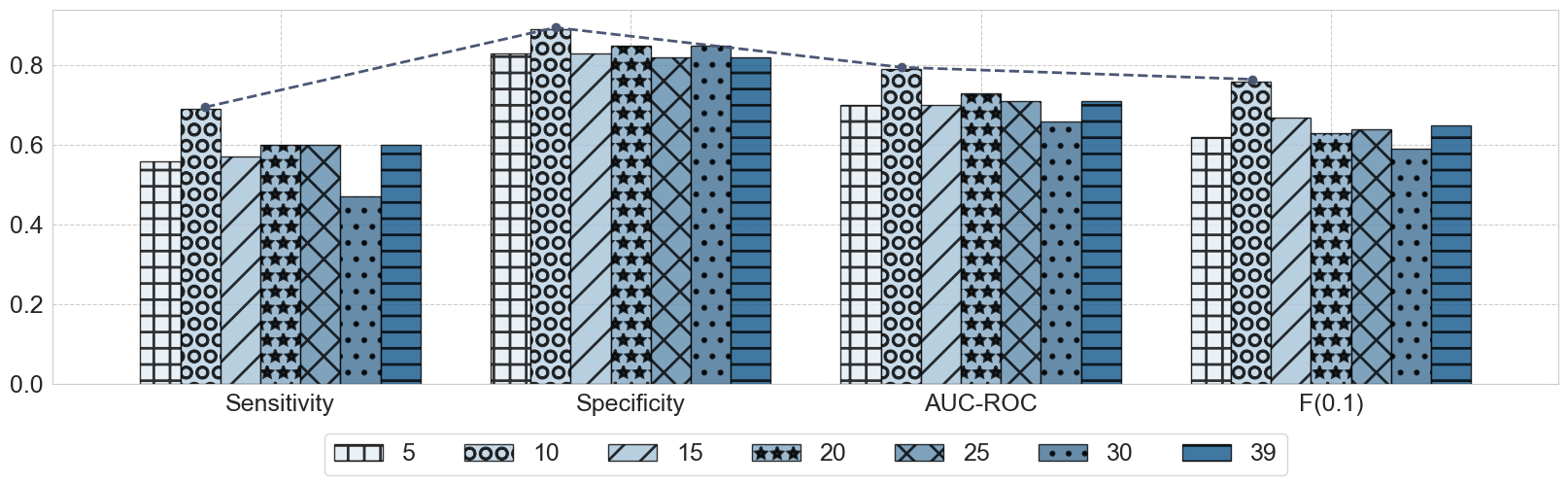}
\caption{Effect of Different Numbers of Sensor Features selected by Mutual Information on the performance of \textit{CovidRhythm} }
\label{fig:statiticalfeatures}
\end{figure*}

In order to achieve the best possible results, it was necessary to select the optimal number of features to input to the model. Mutual Information (section \ref{featureselection}) was utilized to reduce the dimensions of the feature space. Result in Figure \ref{fig:statiticalfeatures} show that the optimal performance was achieved when 10 sensor features were used. Subsequently, we utilized only 10 rhythmic features for training \textit{CovidRhythm} .

\subsection{Contribution of Rhythmic Features}
To drill down further into the importance of biobehavioral rhythmic features for passive identification of Covid-19, we trained \textit{CovidRhythm}  and baseline models on combinations of sensor and rhythmic features (Table \ref{tab:sensorvsrhythm}). We observed that rhythmic features significantly impacted performance when used alone or in combination with sensor features. All classifier types could accurately use sensor data to  identify healthy subjects; however, they fail to predict Covid-19 infected samples. However, rhythmic features were more predictive of infection. After fusing the sensor data with rhythmic features, CovidRhythm outperformed all combinations. When trained on sensor + rhythmic features, the performance of traditional machine learning algorithms dropped due to their inability to analyze time series data. On the other hand, when they were trained only on rhythmic features, their performance was comparable to CovidRhythm. Naive Bayes achieved the highest sensitivity of 0.75 among all the algorithms and combinations; however, it obtained the F$_{0.1}$ of 0.51, indicating a high false positive rate. 

\subsection{Evaluation of the Effects of Different Numbers of Heads}

\begin{figure}[t] 
 \includegraphics[width=\columnwidth]{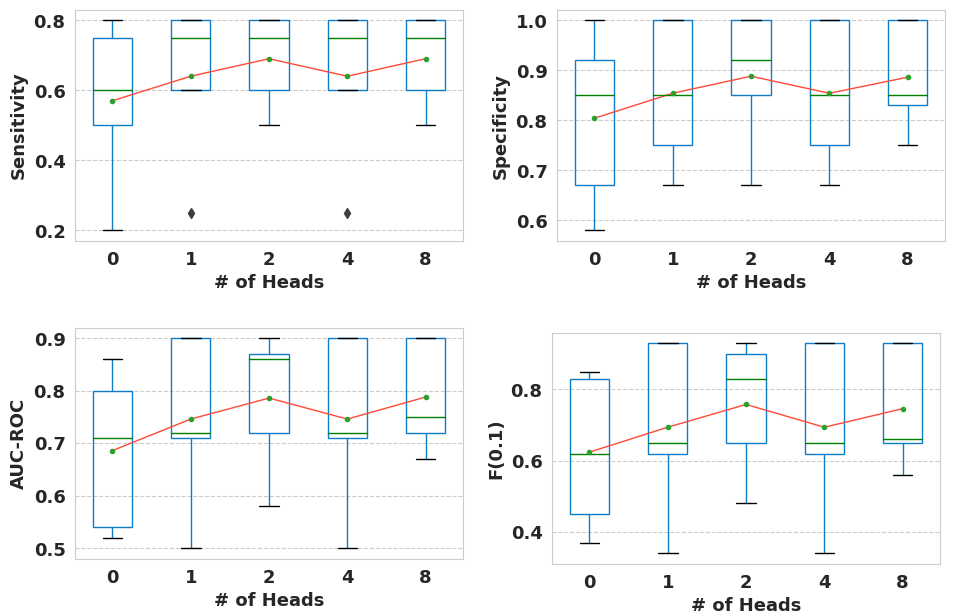}
 \caption{Value of Evaluation Metrics (Sensitivity, Specificity, AUC-ROC, and F$_{0.1}$) for Various Numbers of Heads in Multi-Head Self Attention}
\label{fig:heads}
\end{figure}

In this section, we aimed to evaluate how the different number of heads in MHSA influence the performance of \textit{CovidRhythm}. Figure \ref{fig:heads} showed the results when trained with 0, 1, 2, 4, and 8 heads, where 0 refers to not implementing the multi-head attention.  We found that I) Using no MHSA head produced the worst results, indicating that MHSA was necessary to capture relationships between inputs, II) Optimal performance was achieved when \textit{CovidRhythm}  was trained using two MHSA heads, implying that input data were related with each other in 2 aspects, and III) Beyond two heads, performance reduced indicating that MHSA heads were becoming redundant as no additional relationships could be leveraged between the data inputs. As the data points contribute unequally in making the final predictions from time-series data, determining the best representations for employing MHSA boosted performance.

\subsection{Sub-Sequence Resolution}
\begin{figure}[t]
\centering
\includegraphics[width=\columnwidth]{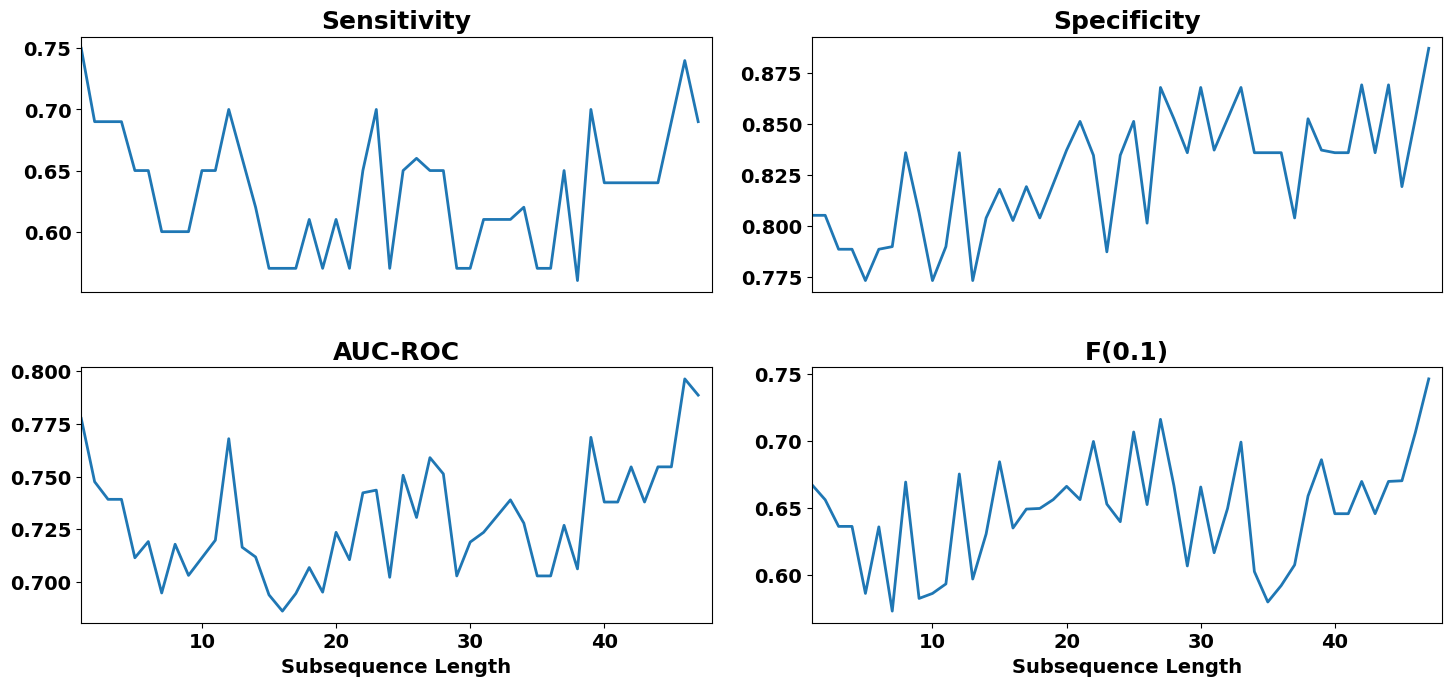}
\caption{Value of Evaluation Metrics (Sensitivity, Specificity, AUC-ROC, and F-Beta(0.1)) for Different Sub-Sequence Lengths }
\label{fig:subsequencelen}
\end{figure}

\textit{CovidRhythm} aims to predict Covid-19 infection from heart rate and steps measured over 24h, segmented into hourly windows with 50\% overlapping, yielding 48 segments. In this analysis, we aim to determine whether using the entire sequence of 48 segments achieves the highest performance or if there exist a sub-sequence length/resolution that can achieve better/same results. We trained \textit{CovidRhythm}  on a time series of different sub-sequence lengths, i.e., $length = 1,2,3,....47$, to determine the optimal length. Figure \ref{fig:subsequencelen} provides the results as a function of different sub-sequence resolutions. We observed that \textit{CovidRhythm}  achieved the highest metric values at the sub-sequence length of 1, 46, and 47. However, even though the performance is promising in terms of sensitivity, specificity, and AUC-ROC at the length of 1, the F$_{0.1}$ is low, suggesting that \textit{CovidRhythm}  is generating too many false positives. We found that the ideal sub-sequence length is either 46 or 47. The selection of appropriate sub-sequence resolution varies greatly with the question: "What is more important - identifying healthy subjects or infected ones?". We selected 47 as the sub-sequence resolution as it has more stable performance in terms of correctly identifying the healthy subjects and F$_{0.1}$.

\subsection{Rhythm Periodicity}

\begin{figure}[t]
\centering
\includegraphics[width=14cm]{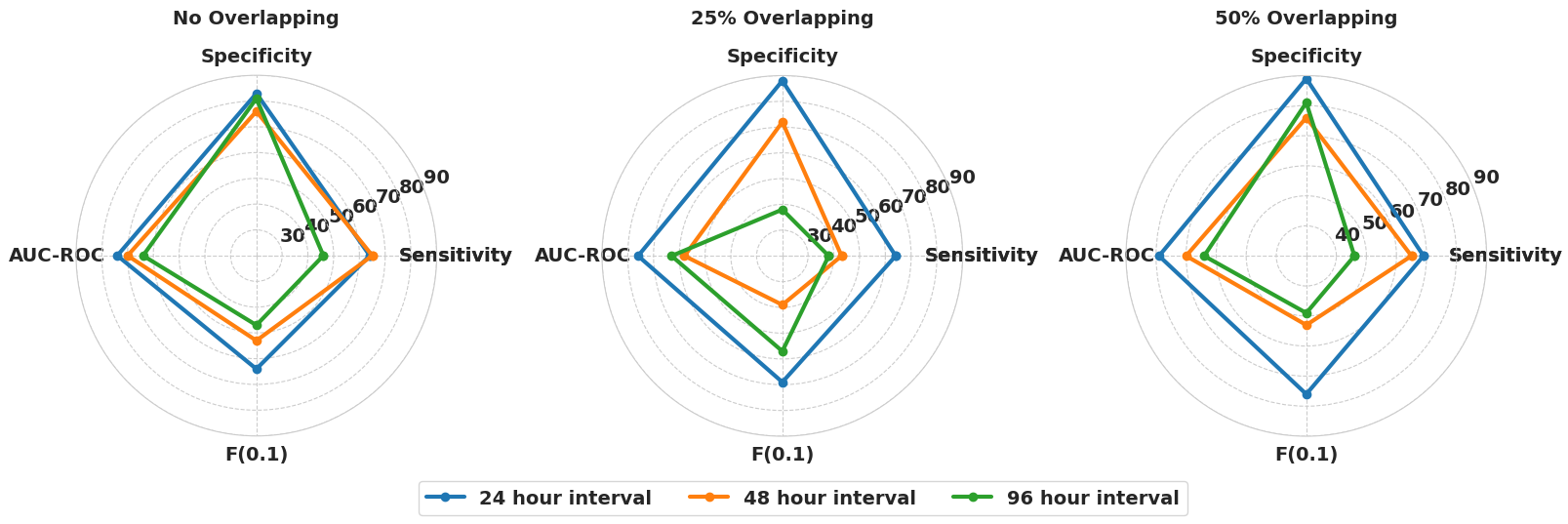}
\caption{Performance of \textit{CovidRhythm}  as a Function of Different Intervals for Measuring Rhythmicity}
\label{fig:periodicity}
\end{figure}

Based on their patterns of reoccurrence, bio-behavioral rhythms are categorized as ultradian, circadian, and infradian. To investigate which rhythms were more disrupted by Covid-19 infection, we trained \textit{CovidRhythm} on features extracted over 24h, 48h, and 96h period. Figure \ref{fig:periodicity} shows the results of \textit{CovidRhythm}  when trained with rhythmic features extracted over three periods. The results show that for Covid-19 patients, the variations in circadian rhythms are the most obvious ones. We speculate that this is because heart rate  and rest-activity/sleep-wake rhythms undergo marked fluctuations over the 24-h day \cite{vandewalle2007robust}\cite{rar} compared to 48hr and 96hr, conveying more meaningful information in terms of rhythm disruption.

 \begin{table*}[!h]
  \caption{Results of \textit{CovidRhythm} vs Baselines on Combinations of Sensor and Rhythmic Features}
  \label{tab:sensorvsrhythm}
  \begin{tabular}{p{2cm}p{0.85cm}p{0.85cm}p{0.85cm}p{1cm}p{0.85cm}p{0.85cm}p{0.85cm}p{1cm}p{0.85cm}p{0.85cm}p{0.85cm}p{0.9cm}}
\hline
 &
\multicolumn{4}{c}{\textbf{Sensor}} &
\multicolumn{4}{c}{\textbf{Rhythmic}} &
\multicolumn{4}{c}{\textbf{Sensor + Rhythmic}}\\
\hline
\textbf{Algo} & \textbf{Sens.} & \textbf{Spec.} &\textbf{AUC}& \textbf{F$_{0.1}$} & \textbf{Sens.} & \textbf{Spec.} & \textbf{AUC} & \textbf{F$_{0.1}$}& \textbf{Sens.} & \textbf{Spec.} &\textbf{AUC}& \textbf{F$_{0.1}$} \\
\hline
Naive Bayes  & 0.44$_{\pm{0.1}}$ & 0.5$_{\pm{0.2}}$ & 0.48$_{\pm{0.1}}$ & 0.26$_{\pm{0.1}}$ &0.75$_{\pm{0.2}}$ &0.7$_{\pm{0.1}}$ &0.8$_{\pm{0.1}}$ &0.51$_{\pm{0.1}}$ & 0.44$_{\pm{0.1}}$ & 0.51$_{\pm{0.2}}$ &0.49$_{\pm{0.2}}$ & 0.27$_{\pm{0.1}}$\\
\hline
Random Forest & 0.17$_{\pm{0.2}}$  & 0.92$_{\pm{0.1}}$  & 0.6$_{\pm{0.1}}$  & 0.4$_{\pm{0.4}}$  &0.66$_{\pm{0.2}}$  &0.85$_{\pm{0.1}}$ & 0.86$_{\pm{0.1}}$&0.7$_{\pm{0.3}}$ &  0.22$_{\pm{0.2}}$ & 0.98$_{\pm{0.1}}$ & 0.77$_{\pm{0.1}}$ & 0.68$_{\pm{0.4}}$\\
\hline
Support Vector Machines   & 0.36$_{\pm{0.3}}$  & 0.9$_{\pm{0.1}}$  & 0.63$_{\pm{0.1}}$  & 0.55$_{\pm{0.4}}$   &0.61$_{\pm{0.2}}$  &0.83$_{\pm{0.2}}$  &0.72$_{\pm{0.1}}$  & 0.64$_{\pm{0.3}}$ &  0.36$_{\pm{0.3}}$ & 0.90$_{\pm{0.1}}$ & 0.63$_{\pm{0.1}}$ & 0.4$_{\pm{0.1}}$ \\
\hline
Gradient Boosting  & 0.21$_{\pm{0.2}}$ & 0.9$_{\pm{0.1}}$ & 0.56$_{\pm{0.1}}$ & 0.31$_{\pm{0.3}}$ & 0.62$_{\pm{0.1}}$  &0.84$_{\pm{0.2}}$ &0.86$_{\pm{0.1}}$ & 0.64$_{\pm{0.2}}$& 0.35$_{\pm{0.2}}$ & 0.82$_{\pm{0.2}}$ & 0.65$_{\pm{0.1}}$ & 0.38$_{\pm{0.3}}$\\
\hline
AdaBoost  & 0.37$_{\pm{0.3}}$ & 0.72$_{\pm{0.2}}$ & 0.54$_{\pm{0.2}}$ & 0.23$_{\pm{0.2}}$ & 0.38$_{\pm{0.3}}$ &0.84$_{\pm{0.1}}$ & 0.74$_{\pm{0.1}}$&0.44$_{\pm{0.3}}$ & 0.59$_{\pm{0.4}}$ & 0.87$_{\pm{0.1}}$ & \textbf{0.87}$_{\pm{0.1}}$ & 0.58$_{\pm{0.4}}$\\
\hline
CNN   & 0.0 & 0.9$_{\pm{0.2}}$ & 0.45$_{\pm{0.1}}$ & 0.0 & 0.08$_{\pm{0.2}}$  & 0.97$_{\pm{0.1}}$ &0.52$_{\pm{0.1}}$  &0.1$_{\pm{0.2}}$  &0.66$_{\pm{0.2}}$ & 0.82$_{\pm{0.2}}$ & 0.74$_{\pm{0.1}}$ & 0.65$_{\pm{0.2}}$ \\
\hline
LSTM  & 0.43$_{\pm{0.3}}$ & 0.78$_{\pm{0.1}}$ & 0.61$_{\pm{0.2}}$ &  0.44$_{\pm{0.3}}$ & 0.55$_{\pm{0.2}}$  &0.77$_{\pm{0.1}}$  &0.66$_{\pm{0.1}}$  &0.47$_{\pm{0.1}}$  &  0.52$_{\pm{0.2}}$ & 0.82$_{\pm{0.2}}$ & 0.68$_{\pm{0.3}}$ & 0.58$_{\pm{0.4}}$ \\
\hline
CNN + LSTM  & 0.34$_{\pm{0.3}}$ & 0.76$_{\pm{0.2}}$ & 0.55$_{\pm{0.2}}$ & 0.64$_{\pm{0.3}}$ &0.51$_{\pm{0.2}}$ &0.71$_{\pm{0.1}}$  & 0.61$_{\pm{0.1}}$ &0.39$_{\pm{0.1}}$  & 0.69$_{\pm{0.1}}$ & 0.80$_{\pm{0.2}}$ & 0.75$_{\pm{0.2}}$ & 0.64$_{\pm{0.3}}$ \\
\hline
\textbf{CovidRhythm}& 0.27$_{\pm{0.3}}$  & 0.77$_{\pm{0.2}}$ & 0.52$_{\pm{0.2}}$ & 0.25$_{\pm{0.2}}$ &0.53$_{\pm{0.3}}$ &0.75$_{\pm{0.2}}$  &0.64$_{\pm{0.2}}$ &0.55$_{\pm{0.3}}$  & \textbf{0.69}$_{\pm{0.1}}$ & \textbf{0.89}$_{\pm{0.1}}$ & 0.79$_{\pm{0.1}}$ & \textbf{0.76}$_{\pm{0.2}}$ \\
 \hline
\end{tabular}
\end{table*}


\section{Discussion}\label{discussion}
\textbf{Main Findings} Consumer-grade wearables have opened up new opportunities for the passive assessment of Covid-19 at scale. Even though the reliability of these wearables compared to clinical-grade wearables, has always been questioned, the recent literature demonstrates that they can be a valuable resource for analyzing bio-rhythms. This study analyzed the biobehavioral rhythms extracted from heart rate and steps measurements captured using Fitbit to predict Covid-19 infection.

\textbf{Rhythmic Features were predictive of Covid-19 infection}. In contrast, sensor features (RHR Mean/Entropy/Slope, Max/Min positive change in RHR, Max. no. of steps taken in 5-min interval, Max steps taken in active bout) were more effective in identifying healthy subjects. We speculate that this was probably linked with the low Recall (Sensitivity) achieved by prior  studies in which one type of sensor data (heart rate, and steps) was collected passively in the wild and employed to predict Covid-19 infection. When sensor and rhythmic features were combined and used for training \textit{CovidRhythm}, the sensor features identified healthy subjects accurately, while rhythmic features (M10 for RHR Skewness and Max Positive Change, IV for Total Steps taken and average Steps taken) contributed in predicting Covid-19 cases, jointly achieving an F$_{0.1}$ score of 0.76, sensitivity of 0.69, and specificity of 0.89.

\textbf{Covid-19 infection disrupts Rest-Activity Rhythms the most.} As computed using Mutual Information, the top three most important rhythmic features were related to physical activity (steps), i.e. L5 and MESOR of steps slope, and IV of steps STD. This captures the fact that infected people get less physically active and exhibit a more irregular sleep schedule, disrupting their rest-activity rhythms. These disruptions are the most prominent ones and predict Covid-19 more effectively than abnormalities in heart rate rhythms.

\textbf{Circadian Rhythms were the most disrupted rhythms by Covid-19.} The recurrence of rhythms can range from less or greater than 24 hours up to days, months, and even years. Our experiments revealed that lack of rhythms seen during the 24-hour period was the most accurate for distinguishing the subjects with Covid-19 from healthy controls. This may be because heart rate and rest-activity/sleep-wake rhythms undergo marked fluctuations over the 24-h day, thus the circadian rhythm conveys more meaningful information in terms of rhythm disruption.


\textbf{Deep learning outperformed traditional machine learning techniques}. In our analysis, we found that I) Due to MHSA's ability to compute mulitple weighted representations relevant to the final prediction, employing MHSA boosted performance, increasing F$_{0.1}$ from 0.62 to 0.76, and AUC-ROC from 0.69 to 0.79 when none vs. two heads were utilized. II) Machine learning techniques performed well on rhythmic features, however, they did not perform as well when applied on rhythmic+ sensor features due to their inability to analyze time series data III) Even in cases where the performance of traditional machine learning is comparable to that of deep learning, the former achieved a low F$_{0.1}$ score depicting a high false positive rate.

\textbf{Study limitations} Despite the encouraging results and findings, our work had a few limitations. First, all subjects were Fitbit users and may not be truly representative of the general population. Second, symptom-onset dates were self-reported, and may be inaccurate and effect results. Third, although we utilized a relatively large dataset, the power and generalizability of our analysis were constrained by the dataset size, presenting a challenge for the model to accurately learn baseline behaviors for all subjects. Lastly, other non-Covid factors and conditions such as influenza, chronic pain, and stress can also confound biobehavioral rhythms. To strengthen our findings, additional research with a more diverse dataset is required to distinguish behavioral rhythmic disruptions caused by Covid from other non-Covid factors.

\textbf{Future work} In the future, we aim to investigate the following ideas: I) Are our results confirmed in a larger, more diverse dataset that includes Covid-19 positive patients, healthy controls as well as subjects with infections other than Covid-19? II) Are the results improved if we employ techniques that are not data-hungry and can effectively learn from smaller datasets? III) Can the data from other types and models of wearables reproduce the same results? IV) Apart from steps and heart rate data, how well does other modalities such as sleep, body temperature, and respiration rate perform?

\section{Conclusion}\label{conclusion}

This paper proposed \textit{CovidRhythm}, a deep multi-head self-attention-based gated recurrent unit network to predict Covid-19 from biobehavioral rhythmic features extracted from physiological data captured using Fitbit wearables. We extracted 39 features from sensor data (heart rate and steps) and combined them with 351 biobehavioral rhythmic features, to predict Covid-19 during the incubation period (one day before symptom onset). \textit{CovidRhythm}  achieved an AUC-ROC of 0.79,  sensitivity of 0.69, specificity of 0.89, and F$_{0.1}$  of 0.76. Variations in biobehavioral rhythms, including rest-activity and physiological (heart rate) rhythms performed  best in identifying Covid-19 positive cases using physiological data collected passively in-the-wild.  Rhythmic features had the most discriminating power either used alone or in conjunction with the sensor features. Rhythmic features were more predictive of infection, while the sensor features were more predictive of healthy subjects. Circadian rest-activity rhythms that combine 24h activity and sleep information were the most disrupted. In future, using a larger, more diverse dataset collected from more types of wearables, we believe \textit{CovidRhythm} can be utilized for passive assessment of Covid-19 and ultimately, reducing its spread.

\flushend
\bibliographystyle{unsrt}
\bibliography{main}
\end{document}